\documentclass[11pt,a4paper]{article}
\usepackage[linesnumbered,ruled,vlined]{algorithm2e}
\usepackage{amssymb, amsmath, amsthm}
\usepackage{graphics, color}
\usepackage{multirow}
\usepackage{tikz}
\usepackage{multicol}

\title{Online and semi-online scheduling on two hierarchical machines with a common due date to maximize the total early work}

\author{Man Xiao $^{1}$, Xiaoqiao Liu $^{1}$, Weidong Li $^{1}$,
Xin Chen $^{2,}$\thanks{Corresponding author. e-mail: chenxin.lut@hotmail.com (X. Chen)},\\ Malgorzata Sterna $^{3}$, Jacek Blazewicz $^{3,4}$\\
{\scriptsize $^{1}$ Yunnan University, School of Mathematics and Statistics, Kunming, PR China.} \\
{\scriptsize $^{2}$ Liaoning University of Technology, School of Electronics and Information Engineering, Jinzhou, PR China.}\\
{\scriptsize $^{3}$ Poznan University of Technology, Institute of Computing Science, Poznan, Poland.}\\
{\scriptsize $^{4}$ European Centre for Bioinformatics and Genomics, Polish Academy of Sciences, Poznan, Poland.}\\
}
\date{}

\begin{document}

\maketitle

\begin{abstract}
\noindent In this study, we investigated several online and semi-online scheduling problems on two hierarchical machines
with a common due date to maximize the total early work.
For the pure online case, we designed an optimal online algorithm with a competitive ratio of $\sqrt 2$.
For the case when the total processing time is known, we proposed an optimal semi-online algorithm with a competitive ratio of $\frac{4}{3}$.
Additionally, for the cases when the largest processing time is known, we gave optimal algorithms
with a competitive ratio of $\frac{6}{5}$ if the largest job is a lower hierarchy one,
and of $\sqrt 5-1$ if the largest job is a higher hierarchy one, respectively.\\

\end{abstract}

\section{Introduction}
Early work scheduling on parallel machines with a common due date \cite{Chen2016} involves scheduling $n$ jobs on $m$ parallel identical machines in a non-overlapping and non-preemptive manner, to maximize the total early work of all the jobs. The early work of a job denotes its part executed before the common due date.
From the optimization point of view, early work maximization is correlated to late work minimization, since the late work of a job denotes its part executed after this due date. Both measures have been widely investigated  over several years \cite{Sterna2011, Sterna2021} from a theoretical perspective as well as in various practical applications,
such as in control systems when collecting data from sensors, in agriculture in the process of harvesting crops,
in manufacturing systems in planning technological processes, and in software engineering in the process of software testing. Late and early work are obviously equivalent when the optimal solutions should be determined. But the studies on offline approximation and online algorithms have been done mostly for early work.

For an offline maximization problem, a $\rho$-approximation algorithm $A$ is a polynomial time algorithm that always obtains a feasible solution (for any instance $I$ with a minimum $\rho$) satisfying $\frac{C^{OPT}(I)}{C^A(I)} \leq \rho$, in which $C^{OPT}(I)$($C^{OPT}$ for short) denotes the optimal criterion value, and $C^A(I)$($C^A$ for short) denotes the output value by $A$. In case of offline scheduling, the best approximation methods are approximation schemes, which allow to control the quality of obtained solutions.
A polynomial time approximation scheme (PTAS) for a given maximization problem is a
family of approximation algorithms which guarantee that $C^A \geq (1-\epsilon)C^{OPT}$ for any $\epsilon \in (0,1)$.
An efficient polynomial time approximation scheme (EPTAS) is a PTAS with the running time bounded by a value of $f(\frac{1}{\epsilon})poly(|I|)$, where $f$ is a computable function but not necessarily polynomial, and $poly(|I|)$ is
a polynomial of the length of the (binary) encoding of the input.
A fully polynomial time approximation scheme (FPTAS) is an EPTAS which satisfies that $f$ is polynomial in $\frac{1}{\epsilon}$.

In this paper, we study online scheduling problems where the set of jobs is unknown in advance and jobs arrive one by one. The lack of knowledge on the problem instance limits the optimization process, but there is possible to construct online algorithms, which efficiency is evaluated in a similar way to the above mentioned offline approximation algorithms.

The performance of an online algorithm is measured by its competitive ratio. Similar with the approximation ratio, the competitive ratio of an online algorithm $A$ is defined as the minimum $r$ such that $\frac{C^{OPT}}{C^A} \leq r$ holds for any problem instance. Then we claim that the considered scheduling problem has an upper bound of $r$.
On the other hand, proving that no online algorithm has a competitive ratio less than $\delta$ denotes that $\delta$ is a lower bound of this problem.
In consequence, an online algorithm is called optimal (i.e., the best possible) if its competitive ratio $r$ equals to the problem lower bound $\delta$, i.e., $r=\delta$.
In such a case, we call that this problem has a tight bound.

Previous studies on early work maximization for parallel identical machines and a common due date focus mainly on offline scheduling. The non-preemptive problem is weakly NP-hard for the fixed number of  machines \cite{Chen2016}, while for an arbitrary number of machines it becomes strongly NP-hard \cite{Chen2020b}. For the two-machine model ($m=2$), Sterna and Czerniachowska \cite{Sterna2017} designed a PTAS based on structuring the problem input. Chen et al. \cite{Chen2020b} proved that the classical LPT (largest processing time first) heuristic has an approximation ratio of $\frac{10}{9}$, and then, Jiang et al. \cite{Jiang2021} demonstrated that the tight bound of LPT is exactly $\frac{12}{11}$.
Then, for the more general case where the number of machine $m$ is fixed ($m \geq 2$), Chen et al. \cite{Chen2020} proposed an FPTAS based on a dynamic programming approach.
Finally, for the cases with is an arbitrary number of machines, Gy\"{o}rgyi and Kis \cite{Gyorgyi2020} proposed a PTAS, while Li \cite{Li2020} proposed an EPTAS.

The total early work maximizationn has been introduced into online scheduling by Chen et al. \cite{Chen2016}, who considered the parallel machine model  with a common due date. In particular, they designed an optimal online algorithm with a competitive ratio of $\sqrt 5-1$ for two identical machines. Then, Chen et al. \cite{Chen2021} investigated several semi-online models, where some partial knowledge on problem instances is available in advance. For the cases when the total processing time or optimal criterion value is known, they designed an optimal online algorithm with a competitive ratio of $\frac{6}{5}$.
If the maximal job processing time is known, they set a lower bound of $1.1231$ and an upper bound of $1.1375$, respectively.

In this paper we study the online scheduling problem with a common due date and the total early work taking into account the hierarchical constraint \cite{Bar-Noy2001}. In this model jobs are assigned a certain hierarchy which determines machines on which they can be processed. For the two machine problem, jobs may have lower or higher hierarchy. One machine can process both type of jobs, while the other is dedicated for high hierarchy jobs only.
Hierarchical scheduling has been combined with early work maximization in our previous work \cite{Xiao2021} for the first time. There were studied three semi-online models assuming that the information on jobs' total processing time is provided in advance.
Specifically, when the total processing time of the lower or higher hierarchy is known, the tight bound was proven to be $\sqrt{5} -1$. Additionally, if both of the information are known, the tight bound was shown to be $\frac{6}{5}$.

Following our previous work \cite{Xiao2021}, we consider several - more general - online and semi-online models for early work maximization with the hierarchical constraint and a common due date, and propose efficient approaches for them.
First, we consider the pure online version without any information on job sequence, and design an optimal algorithm with a competitive ratio of $\sqrt 2$.
For the case when the total processing time is known in advance, we propose an optimal semi-online algorithm with a competitive ratio of $\frac{4}{3}$. (Note that this model differs from the ones considered in \cite{Xiao2021} since in the latter we need to know exactly the total processing time of the lower or higher hierarchy, or both of them.)
Finally, for the cases when the maximal job processing time is known,
we design optimal semi-online algorithms with competitive ratios of $\frac{6}{5}$ and $\sqrt 5-1$,
depending on whether the largest job is of a lower hierarchy or a higher hierarchy, respectively.

The rest of this paper is organized as follows.
In Section 2, we give the formal definition for the considered problem.
Then, the pure online case is analysed in Section 3,
followed by the analysis of two semi-online cases in Section 4 and 5, respectively.
Finally, we present conclusions and possible directions for future work in Section 6.

\section{Problem definition}

The two hierarchical machine online scheduling problem with a common due date and the maximum total early work is defined as follows.

There are two parallel machines $M=\{M_1, M_2\}$, and a set of jobs $J={\{J_1, J_2, ..., J_n}\}$, which come to the system one by one. We say that these jobs come ``over a list" in contrary to online problems where jobs arrive at given time moments. Once a job arrives, it has to be scheduled immediately and irrevocably on one of the machines. The $j$-th job $J_j$ is described by two parameters $(p_j, g_j)$, where $p_j$ is the processing time and $g_j\in{\{1,2}\}$ is the hierarchy of this job.
The hierarchical constraint means that the machines have different capabilities when processing jobs. $M_1$ is a general machine which is available for all the jobs, while $M_2$ is a specific one, available only for the high-hierarchy jobs (such as VIP service in a bank system). Hence, $M_1$ can process all the jobs (with $g_j\in{\{1,2}\}$), whereas $M_2$ can process only jobs of hierarchy 2 ($g_j=2$).
Moreover, in the common due date models considered in this paper, all the jobs share a common due date $d>0$ ($d_j=d$), before which they should be preferably scheduled. We can assume that $p_{j}\leq d$ for $j=1,2,\ldots,n$.

The goal is to schedule jobs from $J$ on machines from $M$ without preemption to maximize the total early work of all jobs. The early work of job $J_j$, denoted as $X_j$, is the part of $J_j$ executed before the common due date $d$, i.e., $X_j = \min\{p_j, \max\{0, d-(C_j-p_j)\}\}$, where $C_j$ denotes the job completion time. More precisely, if $J_j$ is completed before $d$ ($C_j\leq d$), this job is called totally early and we have $X_j=p_j$. If $J_j$ starts before $d$, ($C_j-p_j<d$), but finishes its execution after $d$ ($C_j>d$), we say that this job is partially early (or partially late) and $X_j=d-(C_j-p_j)$.
Finally, if $J_j$ starts its execution after $d$ ($C_j-p_j\geq d$), this job is totally late and we have no profit in this case, i.e., $X_j=0$.

Using the three-filed notation  commonly applied in the scheduling domain \cite{Graham1979}, the pure online model of the considered problem can be denoted as $P2|GoS,online,d_j=d|\max(X)$, in which $GoS$ denotes the hierarchical constraint.

Taking into account the definition of the considered problem, we see  that a schedule of $J$ on $M$ can be considered as a partition $(S_1,S_2)$ of all jobs to two machines, such that $S_1 \cup S_2=J$ and $S_1\cap S_2=\varnothing$.
We will define the load of machine $M_i$ ($i\in{\{1,2}\}$) by $L_i$, i.e., $L_i=\sum_{J_j\in{S_i}}p_j$. Then the early work of jobs assigned to $M_i$ is equal to $\min\{L_i,d\}$. Thus, the aim of the scheduling is to find a partition such that $X=\sum_{j=1}^nX_j=\sum_{i=1}^{2}\min{\{L_i,d}\}$ is maximized.

Let $L^j_i$ be the load of $M_i$ after job $J_j$ is assigned to one of the machines ($i \in \{1,2\}$ and $1 \leq j \leq n$), so we have $L_i = L^n_i$ for each machine. Moreover, let $T_1$ or $T_2$ be the total processing time of the jobs with hierarchy 1 or 2, and let $T$ be the total processing time of all jobs, i.e., $T = T_1+T_2$. Based on the definition of $X$, we have the following lemma. \\

\noindent {\bf Lemma~1.}
In the problem $P2|GoS,online,d_j=d|\max(X)$, the optimal criterion value $C^{OPT}$ satisfies that
\begin{eqnarray*}
C^{OPT}\leq \min{\{T,2d}\} \leq d+\frac{T}{2}.
\end{eqnarray*}

The efficiency of online algorithms proposed for the above defined problem depends on the amount of information on problem instances provided in advance. In the pure online case, $P2|GoS,online,d_j=d|\max(X)$, no additional knowledge is given (cf. Section 3). In the semi-online cases some pieces of information are given in advance. We will consider, the problems where the total job processing time is known, $P2|GoS,online,d_j=d, T|\max(X)$ (cf. Section 4) and the maximum processing time is known ($p_{max}$), as well as the hierarchy of the job possessing this maximum is known ($p_{max,1}, p_{max,2}$), $P2|GoS,online,d_j=d,\Delta|\max(X)$ (cf. Section 5), where $\Delta \in \{p_{max}, p_{max,1}, p_{max,2}\}$.

\section{Pure online case}
In this section, we study the online problem $P2|GoS,online,d_j=d|\max(X)$, where no information on problem instances is available a priori.
We will prove a lower bound of $\sqrt 2$ for this problem, and provide an optimal online algorithm solving it with a competitive ratio of $\sqrt 2$. It is worth to be mentioned that this model has not been studied before, particularly in our previous research on early work maximization with hierarchical machines \cite{Xiao2021}.\\

\noindent {\bf Theorem~2.} Any online algorithm $A$ for the problem $P2|GoS,online,d_j=d|\max(X)$ has a competitive ratio of at least $\sqrt 2$. \\

\noindent  {\bf Proof.} Let $d=1$. The first job in the job sequence is $J_1=(\sqrt 2-1,2)$.

If $J_1$ is assigned to $M_1$, then the last job is $J_2=(1,1)$, which implies that $C^{OPT}=\sqrt 2$, and $C^A=1$. 

If $J_1$ is assigned to $M_2$, the second job $J_2=(1,2)$ arrives. If $J_2$ is assigned to $M_2$, no more jobs arrive, which implies that $C^{OPT}=\sqrt{2}$ and $C^A=1$. If $J_2$ is assigned to $M_1$, the last job is $J_3=(1,1)$. Hence, we have $C^{OPT}=2$, which is obtained by assigning  $J_2,J_3$ to different machines and $C^A=\sqrt 2$.

Therefore, in any case, we have $\frac{C^{OPT}}{C^A}\geq\sqrt 2$, implying that the theorem holds. \qed \\

Now we will propose an optimal online algorithm solving the considered pure online scheduling problem, Algorithm 1. Its primary concept is that large jobs are not assigned to $M_2$ (the machine dedicated for high hierarchy jobs only) unless the current load of $M_2$ is relatively small.\\

  \begin{algorithm}[!htbp]  \label{A1}
 	\caption{}
 	\LinesNumbered
 	
 	Initially, let $L^0_2=0$.
 	
 	When a new job $J_{j}=(p_j,g_j)$ arrives,
 	
 	\If{$g_j=1$}{Assign job $J_j$ to $M_1$, and let $L^{j}_2=L^{j-1}_2$.}
 	\Else {

         \If{$L^{j-1}_2+p_j\leq d$}{Assign $J_j$ to $M_2$, and set $L^{j}_2=L^{j-1}_2+p_j$.}
         \Else {
               \If{$L^{j-1}_2\leq(\sqrt{2}-1)d$}{Assign $J_j$ to $M_2$, and set $L^{j}_2=L^{j-1}_2+p_j$.}
               \Else{Assign $J_j$ to $M_1$, and set $L^{j}_2=L^{j-1}_2$.}
               }
 	      }
 	
 	If there is another job, $j:=j+1$, go to {\bf step 2.}
 	Otherwise, stop.
 \end{algorithm}

\noindent {\bf Theorem~3.} The competitive ratio of Algorithm 1 is at most $\sqrt 2$.\\

\noindent  {\bf Proof.} Based on Lemma 1, if $\min{\{L_1,L_2}\}\geq d$, we have $C^A=2d\geq C^{OPT}$.
If $\max{\{L_1,L_2}\}\leq d$, we have $C^A=T\geq C^{OPT}$. It implies that we only need to consider the case when
 $\min{\{L_1,L_2}\}<d<\max{\{L_1,L_2}\}$, implying that
  \begin{eqnarray*}
  C^A=d+\min{\{L_1,L_2}\}.
  \end{eqnarray*}
Subsequently, we distinguish the following two cases.

\noindent {\bf Case 1.} $\max{\{L_1,L_2}\}=L_1>d$

In this case, we have $C^A=d+L_2$.
If there is no job of hierarchy 2 assigned to $M_1$, we have
$L_1=T_1>d$ and $L_2=T_2<d$, implying that Algorithm 1 reaches the optimal solution.
Else, let $J_l=(p_l,2)$ be the last job of hierarchy 2 assigned to $M_1$. By the choice
of Algorithm 1, we have
 $L_2\geq L^{l-1}_2>(\sqrt 2-1)d$. By Lemma 1, we have
\begin{eqnarray*}
\frac{C^{OPT}}{C^A}\leq\frac{2d}{d+L_2}\leq\frac{2d}{d+(\sqrt 2-1)d}=\sqrt 2.
\end{eqnarray*}

\noindent {\bf Case 2.} $\max{\{L_1,L_2}\}=L_2>d$

In this case, we have $C^A=d+L_1$. Let $L_{1,2}$ be the total processing time of jobs of hierarchy 2 assigned to $M_1$, and $J_l=(p_l,2)$ be the last job assigned to $M_2$.

If $J_l$ is assigned to $M_2$ at Line 7 of the algorithm, we have $L_2=L^l_2=L^{l-1}_2+p_l\leq d$, contradicting
the assumption $L_2>d$.

If $J_l$ is assigned to $M_2$ at Line 10, we have $L_2=L^l_2=L^{l-1}_2+p_l>d$ and $L^{l-1}_2\leq(\sqrt 2-1)d$. Therefore,
$T_2=L^{l-1}_2+p_l+L_{1,2}\leq (\sqrt 2-1)d+p_l+L_{1,2}\leq\sqrt {2}d+L_{1,2}$,
where the last inequality follows from the assumption $p_l \leq d$. Based on Lemma 1, we have
\begin{eqnarray*}
\frac{C^{OPT}}{C^A}\leq\frac{T}{d+L_1}=\frac{T_1+T_2}{d+T_1+L_{1,2}}\leq\frac{T_1+\sqrt 2d+L_{1,2}}{d+T_1+L_{1,2}}\leq\sqrt 2.
\end{eqnarray*} \qed

\section{Semi-online case with total processing time known}

In this section, we provide the details for the design of an optimal semi-online algorithm for the case when the total processing time of all jobs ($T$) is known in advance. This problem is denoted as $P2|GoS,online,d_j=d,T|\max(X)$. For this case, we prove a lower bound of $\frac{4}{3}$, and design an algorithm with a competitive ratio of exactly $\frac{4}{3}$.\\

\noindent {\bf Theorem~4.} Any online algorithm $A$ for the problem $P2|GoS,online,d_j=d,T|\max(X)$ has a competitive ratio of at least $\frac{4}{3}$. \\

\noindent  {\bf Proof.} Let $d=1$ and $T=2$. The first job is $J_1=(\frac{1}{2},2)$.

If $J_1$ is assigned to $M_1$, the last two jobs $J_2=(1,1)$ and $J_3=(\frac{1}{2},2)$ arrive. Therefore, $C^{OPT}=2$, and $C^A\leq \frac{3}{2}$.

If $J_1$ is assigned to $M_2$, the next job $J_2=(1,2)$ arrives. If $J_2$ is assigned to $M_1$, the last job $J_3=(\frac{1}{2},1)$ arrives. Therefore, $C^{OPT}=2$ and $C^A=\frac{3}{2}$. If $J_2$ is assigned to $M_2$, the last job $J_3=(\frac{1}{2},2)$ arrives. Therefore, $C^{OPT}=2$ and $C^A\leq\frac{3}{2}$.

Hence, in any case, we have $\frac{C^{OPT}}{C^A}\geq\frac{4}{3}$. \qed \\

The semi-online algorithm, Algorithm 2, solving the considered problem is presented below. Because $M_1$ is capable of processing all jobs, our approach is to first allocate a certain amount of jobs of hierarchy 2 to $M_2$, and subsequently assign the remaining jobs to machine $M_1$. The competitive ratio for this approach is proved in Theorem 5.\\

\begin{algorithm}[!htbp]  \label{A2}
 	\caption{}
 	\LinesNumbered
 	
 	Initially, let $L^0_1=L^0_2=0$.
 	
 	When a new job $J_{j}=(p_j,g_j)$ arrives,
 	
 	\If{$g_j=1$}{Assign the job $J_j$ to $M_1$, and update $L^{j}_1, L^{j}_2$.}
 	\Else {

         \If{$T-L^{j-1}_2-p_j>\frac{5}{8}T$}{Assign the $J_j$ to $M_2$, and update  $L^{j}_1, L^{j}_2$.}
         \Else {
               \If{$T-L^{j-1}_2-p_j\geq L^{j-1}_2$}{Assign the $J_j$ to $M_2$, and assign the remaining jobs to $M_1$(if there are jobs after $J_j$).}
               \Else{Assign the $J_j$ to $M_1$, and assign the remaining jobs to $M_1$(if there are jobs after $J_j$). Update $L^{j}_1, L^{j}_2$.}
               }
 	      }
 	
 	If there is another job, $j:=j+1$, go to {\bf step 2.}
 	Otherwise, stop.
 \end{algorithm}

\noindent {\bf Theorem~5.} The competitive ratio of Algorithm 2 is at most $\frac{4}{3}$.\\

\noindent  {\bf Proof.} If $\min{\{L_1,L_2}\}\geq d$ or $\max{\{L_1,L_2}\}\leq d$,
Algorithm 2 reaches the optimal partition of jobs to machines. We only need to consider the case when
 $\min{\{L_1,L_2}\}<d<\max{\{L_1,L_2}\}$, which implies that
  \begin{eqnarray*}
  C^A=d+\min{\{L_1,L_2}\}.
  \end{eqnarray*}
Subsequently, we distinguish the following two cases:

\noindent {\bf Case 1.} $\max{\{L_1,L_2}\}=L_1>d$

In this case, we have $C^A=d+L_2$. If there is no job of hierarchy 2 assigned to $M_1$, we have
$L_1=T_1>d$ and $L_2=T_2<d$, implying that Algorithm 2 reaches the optimal solution. Else, let $J_l=(p_l,2)$ be the last job of hierarchy 2 assigned to $M_1$.

If $J_l$ is assigned to machine $M_1$ in Line 10, let $J_{l_2}$ be the last job assigned to $M_2$. By the choice of
Algorithm 2, we have $L_1=T-L^{l_2-1}_2-p_{l_2}\leq \frac{5T}{8}$,
which implies that $L_2\geq\frac{3T}{8}$.
Based on Lemma 1, we have
\begin{eqnarray*}
\frac{C^{OPT}}{C^A}\leq\frac{d+\frac{T}{2}}{d+L_2}\leq\frac{d+\frac{T}{2}}{d+\frac{3T}{8}}
=1+\frac{\frac{T}{8}}{d+\frac{3T}{8}}\leq\frac{4}{3}.
\end{eqnarray*}

If $J_l$ is assigned to $M_1$ in Line 12, let $J_{l_1}$ be the first job assigned to $M_1$ in Line 12. By the choice of
Algorithm 2, we have $L_2=L^{l_1-1}_2> T-L^{l_1-1}_2-p_{l_1}=L_1-p_{l_1}$, implying that $L_2>\frac{T-p_{l_1}}{2}$. If $T\geq 2d$, based on Lemma 1, we have
\begin{eqnarray*}
\frac{C^{OPT}}{C^A}\leq\frac{d+\frac{T}{2}}{d+L_2}\leq\frac{d+\frac{T}{2}}{d+\frac{T-p_{l_1}}{2}}
\leq\frac{d+\frac{T}{2}}{\frac{T+d}{2}}=\frac{2d+T}{T+d}=1+\frac{d}{T+d}\leq 1+\frac{d}{3d}=\frac{4}{3},
\end{eqnarray*}
where the third inequality follows from the assumption $p_{l_1}\leq d$. If $T<2d$, we have
\begin{eqnarray*}
\frac{C^{OPT}}{C^A}\leq\frac{T}{d+L_2}\leq\frac{T}{d+\frac{T-p_{l_1}}{2}}\leq\frac{T}{\frac{T+d}{2}}=\frac{2T}{T+d}
\leq\frac{2T}{\frac{3T}{2}}=\frac{4}{3}.
\end{eqnarray*}

\noindent {\bf Case 2.} $\max{\{L_1,L_2}\}=L_2>d$

In this case, we have $C^A=d+L_1$. Let $L_{1,2}$ be the total processing time of jobs of hierarchy 2 assigned to $M_1$, and $J_l=(p_l,2)$ be the last job assigned to $M_2$.

If $J_l$ is assigned to $M_2$ at Line 7, we have $L_1=T-L^{l-1}_2-p_l>\frac{5T}{8}>L_2$, contradicting
the assumption that $\max{\{L_1,L_2}\}=L_2$.

If $J_l$ is assigned to $M_2$ at Line 10, we have
$L_1=T-L^{l-1}_2-p_l\geq L^{l-1}_2=L_2-p_l$, implying that $L_1\geq\frac{T-p_l}{2}$.
If $T\geq 2d$, we have
\begin{eqnarray*}
\frac{C^{OPT}}{C^A}\leq\frac{d+\frac{T}{2}}{d+L_1}\leq\frac{d+\frac{T}{2}}{d+\frac{T-p_l}{2}}
\leq\frac{d+\frac{T}{2}}{\frac{T+d}{2}}=\frac{2d+T}{T+d}=1+\frac{d}{d+T}\leq 1+\frac{d}{3d}=\frac{4}{3},
\end{eqnarray*}
where the third inequality follows from the assumption $p_{l_1}\leq d$. If $T<2d$, we have
\begin{eqnarray*}
\frac{C^{OPT}}{C^A}\leq\frac{T}{d+L_1}\leq\frac{T}{d+\frac{T-p_l}{2}}\leq\frac{T}{\frac{T+d}{2}}=\frac{2T}{T+d}
\leq\frac{2T}{\frac{3T}{2}}=\frac{4}{3}.
\end{eqnarray*} \qed

It is worth to be mentioned that the considered model, $P2|GoS,online,d_j=d,T|\max(X)$, is not the same as the ones studied in \cite{Xiao2021}, which assume providing more detailed information on the problem instances, such as the total processing time of jobs with particular hierarchies, $T_1$, $T_2$, or both of them.
The new results presented in this section imply that the information of $T$ is useful, since it allows to reduce the tight bound from $\sqrt 2 \approx 1.414$ (in the pure online case, cf. Section 3) to $\frac{4}{3}$ (with the knowledge of $T$).
However, the utility of this piece of information ($T$) is not as helpful as $T_1$ or $T_2$, since the tight bound when knowing the latter ones could be reduced to $\sqrt 5 -1 \approx 1.236$ \cite{Xiao2021}.

\section{Semi-online case with largest processing time known}

In this section, we consider a series of problems when some information on the largest job processing time is known in advance. We denote them as $P2|GoS,online,d_j=d,\Delta|\max(X)$, where $\Delta \in \{p_{max}, p_{max,1}, p_{max,2}\}$.
If $\Delta = p_{max}$, it means that we know the information on the largest processing time of the input jobs. Unfortunately, this information is not useful. Applying the same reasoning as in the proof of Theorem 2, we can get a lower bound of $\sqrt{2}$ for the problem $P2|GoS,online,d_j=d,p_{max}|\max(X)$ with the largest job processing time known (assuming that the largest job processing time is 1). This means that we cannot obtain a better online algorithm with the knowledge of $p_{max}$ compared with the pure online case.

However, the problem lower bound could be reduced if we know more about the job possessing the largest processing time. We define $\Delta = p_{max,k}$ if we know $p_{max}$, and also know that the hierarchy of the job with processing time $p_{max}$ is $k$ ($k \in \{1, 2\}$). For $\Delta = p_{max,1}$, i.e. for the problem $P2|GoS,online,d_j=d,p_{max,1}|\max(X)$, we prove a tight bound $\frac{6}{5}$ (cf. Section 5.1).
In contrast, if the hierarchy of the largest job is 2 ($\Delta=p_{max,2}$), i.e. for the problem  $P2|GoS,online,d_j=d,p_{max,2}|\max(X)$ a tight bound is $\sqrt 5-1$ (cf. Section 5.2).

\subsection{Longest job with low hierarchy}

In this subsection, we focus on the semi-online case when the longest job has hierarchy 1, allowing for its processing on machine $M_1$ only, $\Delta=p_{max,1}$, i.e. for any job $J_j=(p_j,g_j)$:
\begin{eqnarray*}
 p_j\leq p_{max,1}\leq d.
 \end{eqnarray*}
We will prove the lower bound of this problem in Theorem 6 and propose an optimal semi-online algorithm. Since the longest jobs, as any job of hierarchy 1, can be assigned to $M_1$ only, solving the problem we prioritize machine $M_2$ when allocating jobs, which leads to Algorithm 3. Its competitive ratio is proved in Theorem 7.\\

\noindent {\bf Theorem~6.} Any online algorithm $A$ for the problem $P2|GoS,online,d_j=d,p_{max,1}|\max(X)$ has a competitive ratio of at least $\frac{6}{5}$. \\

\noindent  {\bf Proof.} Let $d=1$ and $p_{max,1}=\frac{2}{3}$. The first two jobs are $J_1=(\frac{2}{3},1)$ and $J_2=(\frac{1}{3},2)$. Job $J_1$ can only be assigned to $M_1$.

If $J_2$ is assigned to $M_1$, the last job $J_3=(\frac{2}{3},1)$ arrives, implying that $C^{OPT}=\frac{4}{3}$ and $C^A=1$.

If $J_2$ is assigned to $M_2$, the next job $J_3=(\frac{1}{3},2)$ arrives. Subsequently, we distinguish the following two cases:

\noindent {\bf Case 1.} $J_3$ is assigned to $M_1$.

The last job $J_4=(\frac{2}{3},1)$ arrives, implying that $C^{OPT}=\frac{5}{3}$ and $C^A=\frac{4}{3}$.

\noindent {\bf Case 2.} $J_3$ is assigned to $M_2$.

The next job $J_4=(\frac{2}{3},2)$ arrives. If $J_4$ is assigned to $M_1$,
the last job $J_5=(\frac{2}{3},1)$ arrives, implying that $C^{OPT}=2$ and $C^A=\frac{5}{3}$.
If $J_4$ is assigned to $M_2$, then no more jobs arrive, implying that $C^{OPT}=2$ and $C^A=\frac{5}{3}$.

Thus, $\frac{C^{OPT}}{C^A}\geq\frac{6}{5}$ in any case. \qed \\

\begin{algorithm}[!htbp]  \label{A3}
 	\caption{}
 	\LinesNumbered
 	
 	Initially, let $L^0_2=0$.
 	
 	When a new job $J_{j}=(p_j,g_j)$ arrives,
 	
 	\If{$g_j=1$}{Assign job $J_j$ to $M_1$, and set $L^{j}_2=L^{j-1}_2$.}
 	\Else {

         \If{$L^{j-1}_2<\frac{2d}{3}$}{Assign job $J_j$ to $M_2$, and set $L^{j}_2=L^{j-1}_2+p_j$.}
         \Else {Assign the $J_j$ to $M_1$.}

 	      }
 	
 	If there is another job, $j:=j+1$, go to {\bf step 2.}
 	Otherwise, stop.
 \end{algorithm}

\noindent {\bf Theorem~7.} The competitive ratio of Algorithm 3 is at most $\frac{6}{5}$.\\

\noindent  {\bf Proof.} As before, if $\min{\{L_1,L_2}\}\geq d$ or $\max{\{L_1,L_2}\}\leq d$,
Algorithm 3 reaches the optimal solution. We only need to consider the case when
 $\min{\{L_1,L_2}\}<d<\max{\{L_1,L_2}\}$, which implies that
  \begin{eqnarray*}
  C^A=d+\min{\{L_1,L_2}\}.
  \end{eqnarray*}
We distinguish the following two cases.

\noindent {\bf Case 1.} $\max{\{L_1,L_2}\}=L_1>d$

In this case, we have $C^A=d+L_2$. If there is no job of hierarchy 2 assigned to $M_1$, we have
$L_1=T_1>d$ and $L_2=T_2<d$, which implies that Algorithm 3 reaches the optimal solution.
Else, let $J_l=(p_l,2)$ be the last job of hierarchy 2 assigned to $M_1$.
By the choice of Algorithm 3, we have $L_2\geq L^{l-1}_2\geq\frac{2d}{3}$.
Based on Lemma 1, we have
\begin{eqnarray*}
\frac{C^{OPT}}{C^A}\leq\frac{2d}{d+L_2}\leq\frac{2d}{d+\frac{2d}{3}}=\frac{6}{5}.
\end{eqnarray*}

\noindent {\bf Case 2.} $\max{\{L_1,L_2}\}=L_2>d$

In this case, we have $C^A=d+L_1$. If $p_{max,1}\geq\frac{2d}{3}$, we have $L_1\geq p_{max,1}\geq\frac{2d}{3}$. Therefore,
\begin{eqnarray*}
\frac{C^{OPT}}{C^A}\leq\frac{2d}{d+L_1}\leq\frac{2d}{d+\frac{2d}{3}}=\frac{6}{5}.
\end{eqnarray*}

If $p_{max,1}<\frac{2d}{3}$, let $J_l=(p_l,2)$ be the last job assigned to $M_2$. Assume
that $p_l=\alpha d\leq p_{max,1}$, which implies $\alpha<\frac{2}{3}$. By the choice of Algorithm 3, we have
$L^{l-1}_2<\frac{2d}{3}$ and $L_2=L^l_2=L^{l-1}_2+p_l<\frac{2d}{3}+\alpha d$.
Based on Lemma 1, we have
\begin{eqnarray*}
\frac{C^{OPT}}{C^A}\leq\frac{T}{d+L_1}=\frac{L_1+L_2}{d+L_1}=1+\frac{L_2-d}{L_1+d}
<1+\frac{\frac{2d}{3}+\alpha d-d}{L_1+d}.
\end{eqnarray*}
Since $L_1\geq p_{max,1}\geq p_l=\alpha d$, we have
\begin{eqnarray*}
\frac{C^{OPT}}{C^A}\leq 1+\frac{\frac{2d}{3}+\alpha d-d}{L_1+d}\leq 1+\frac{(\alpha-\frac{1}{3})d}{(1+\alpha)d}
=1+\frac{3\alpha-1}{3+3\alpha}=2-\frac{4}{3+3\alpha}<\frac{6}{5},
\end{eqnarray*}
where the last inequality follows from the fact $\alpha<\frac{2}{3}$. \qed

\subsection{Longest job with high hierarchy}
In this subsection, we focus on the semi-online case when the longest job has hierarchy 2, allowing for its processing on both machines $M_1$ and $M_2$, $\Delta=p_{max,2}$, i.e. for any job $J_j=(p_j,g_j)$:
\begin{eqnarray*}
 p_j\leq p_{max,2}\leq d.
 \end{eqnarray*}

As for the previous case, we will prove the lower bound of this problem in Theorem 8, and propose an optimal algorithm - Algorithm 4. This algorithm reserves machine $M_2$ for the first largest job in the input sequence untill it appears in the system. The optimal competitive ratio of this approach is proved in Theorem 9.\\

\noindent {\bf Theorem~8.} Any online algorithm $A$ for the $P2|GoS,online,d_j=d,p_{max,2}|\max(X)$ has a competitive ratio at least $\sqrt 5-1$. \\

\noindent  {\bf Proof.} Let $d=1$ and $p_{max,2}=\frac{\sqrt 5-1}{2}$. The first job is $J_1=(\frac{\sqrt 5-1}{2},2)$. If $J_1$ is assigned to $M_1$, the last two jobs $J_2=(\frac{1}{2},1)$ and $J_3=(\frac{1}{2},1)$ arrive, which implies that  $C^{OPT}=\frac{\sqrt 5+1}{2}$ and $C^A=1$.

If $J_1$ is assigned to $M_2$, the next job $J_2=(\frac{\sqrt 5-1}{2},2)$ arrives. If $J_2$ is assigned to $M_1$, the last two jobs $J_3=(\frac{1}{2},1)$ and $J_4=(\frac{1}{2},1)$ arrive, which implies that
$C^{OPT}=2$ and $C^A=1+\frac{\sqrt 5-1}{2}=\frac{\sqrt 5+1}{2}$.
If $J_2$ is assigned to $M_2$, no more jobs arrive, which implies that  $C^{OPT}=\sqrt 5-1$ and $C^A=1$.

Thus, $\frac{C^{OPT}}{C^A}\geq \sqrt 5-1$ in any case. \qed  \\

\begin{algorithm}[!htbp]  \label{A4}
 	\caption{}
 	\LinesNumbered
 	
 	Initially, let $L^0_2=0$ and $n_2=0$.
 	
 	When a new job $J_{j}=(p_j,g_j)$ arrives,
 	
 	\If {$g_j=1$}{Assign the job $J_j$ to the $M_1$.}
 	\Else {
          \If {$n_2=0$ \text{ and } $p_j\neq p_{max,2}$}
        {
              \If {$L^{j-1}_2+p_{max,2}+p_j\leq(\sqrt 5-1)d$}{Assign the $J_j$ to $M_2$, and update  $L^j_2$.}
              \Else {Assign the $J_j$ to $M_1$, and update $L^j_2$.}
              }
  \Else {
 	    \If {$n_2=0$ \text{ and } $p_j=p_{max,2}$}{Set $n_2==1$,  assign the $J_j$ to $M_2$, and update  $L^j_2$.}
        \Else {
              \If {$L^{j-1}_2+p_j\leq(\sqrt 5-1)d$}{Assign the $J_j$ to $M_2$, and update  $L^j_2$.}
              \Else {Assign the $J_j$ to $M_1$, and update $L^j_2$.}
       }
 }
 }
 	If there is another job, $j:=j+1$, go to {\bf step 2.}
 	Otherwise, stop.
 \end{algorithm}

\noindent {\bf Theorem~9.} The competitive ratio of Algorithm 4 is at most $\sqrt 5-1$.\\

\noindent  {\bf Proof.} As before, if $\min{\{L_1,L_2}\}\geq d$ or $\max{\{L_1,L_2}\}\leq d$,
Algorithm 4 reaches the optimal solution. We only need to consider the case when
 $\min{\{L_1,L_2}\}<d<\max{\{L_1,L_2}\}$, implying that
  \begin{eqnarray*}
  C^A=d+\min{\{L_1,L_2}\}.
  \end{eqnarray*}
Subsequently, we distinguish the following two cases.

\noindent{\bf Case 1.} $\max{\{L_1,L_2}\}=L_1>d$

In this case, we have $C^A=d+L_2$. If there is no job of hierarchy 2 assigned to $M_1$, we have
$L_1=T_1>d$ and $L_2=T_2<d$, implying that Algorithm 4 reaches the optimal solution.
Else, let $J_l=(p_l,2)$ be the last job of hierarchy 2 assigned to $M_1$, and $J_t=(p_t,2)$ be the
first largest job with $p_t=p_{max,2}$. Clearly, $l\neq t$, as $J_t$ is assigned to $M_2$.

If $l<t$,
by the choice of Algorithm 4, we have
$(\sqrt 5-1)d< L^{l-1}_2+p_{max,2}+p_l\leq 2(L^{l-1}_2+p_{max,2})$. Since $J_t$ is assigned to $M_2$ after job $J_l$,
we have $L_2\geq L^{l-1}_2+p_{max,2}>\frac{(\sqrt 5-1)d}{2}$.

If $l>t$, $J_t$ is assigned to $M_2$ before assigning job $J_l$, which implies that $L^{l-1}_2\geq p_t=p_{max,2}\geq p_l.$
Moreover, by the choice of Algorithm 4, we have
$L^{l-1}_2+p_l>(\sqrt 5-1)d$,
which implies that
$L_2\geq L^{l-1}_2\geq \frac{(\sqrt 5-1)d}{2}$.
Based on Lemma 1, we have
\begin{eqnarray*}
\frac{C^{OPT}}{C^A}\leq\frac{2d}{d+L_2}\leq\frac{2d}{d+\frac{(\sqrt 5-1)d}{2}}=\sqrt 5-1.
\end{eqnarray*}

\noindent{\bf Case 2.} $\max{\{L_1,L_2}\}=L_2>d$.

Obtain $C^A=d+L_1$. Let $J_l=(p_l,2)$ be the last job assigned to $M_2$.
If $J_l$ is assigned to $M_2$ in Line 16, according to the choice of Algorithm 4, $L_2=L^{l-1}_2+p_l\leq (\sqrt 5-1)d$.

If $J_l$ is the first largest job with $p_l=p_{max,2}$, it is assigned to $M_2$ at Line 13. Let $J_t=(p_t,2)$ be the last job assigned to $M_2$ before $J_l$. According to the choice of Algorithm 4, we have $L_2=L^{t-1}_2+p_t+p_l =L^{t-1}_2+p_{max,2}+p_t\leq (\sqrt 5-1)d$.

By Lemma 1, we have
\begin{eqnarray*}
\frac{C^{OPT}}{C^A}\leq\frac{T}{d+L_1}=\frac{L_1+L_2}{d+L_1}\leq\frac{(\sqrt 5-1)d+L_1}{d+L_1}\leq\sqrt 5-1.
\end{eqnarray*}

Therefore, this theorem holds. \qed


\section{Conclusions}
In this paper, we studied several online or semi-online scheduling models with the goal of early work maximization under a common due date in a system consisting of two parallel hierarchical machines. For each model, we proposed an optimal online or semi-online algorithm, by analysing the problem's lower bound and proving the algorithm's competitive ratio.

Particularly, we studied the pure online model where no information on problem input is known in advance, $P2|GoS,online,d_j=d|\max(X)$, the semi-online model where the total processing time of all jobs is provided before scheduling, $P2|GoS,online,d_j=d, T|\max(X)$, as well as the semi-online modesl where the maximum processing time is known together with the hierarchy of the longest job, $P2|GoS,online,d_j=d,p_{max,1}|\max(X)$ and $P2|GoS,online,d_j=d,p_{max,2}|\max(X)$. For these models we designed on optimal (semi) online algorithms with competitive ratios:  $\sqrt 2$,  $\frac{4}{3}$, $\frac{6}{5}$ and $\sqrt 5-1$, respectively.

The obtained results demonstrate that the natural direction for future research could be the analysis of these scheduling models with three or more hierarchical machines.\\

\noindent {\bf Acknowledgements. }  This work is supported in part by the National Natural Science
Foundation of China [Nos. 12071417, 61662088], Program for Excellent
Young Talents of Yunnan University, Training Program of National
Science Fund for Distinguished Young Scholars, IRTSTYN, Key
Joint Project of the Science and Technology Department of Yunnan
Province and Yunnan University [No. 2018FY001(-014)], and by the statutory funds of Poznan University of Technology (Poland).



\end{document}